\documentclass[aps,prl,preprintnumbers,twocolumn,groupedaddress,footinbib]{revtex4-1}

\pdfoutput=1

\usepackage{amsmath,amssymb,mathrsfs}
\usepackage{slashed}
\usepackage{xcolor}
\usepackage[colorlinks=true,linkcolor=blue,citecolor=blue,urlcolor=violet]{hyperref}
\usepackage{tikzfeynman}
\usepackage{soul}
\usepackage{ulem}
\usepackage{youngtab}

\newcommand{\be}{\begin{equation}}
\newcommand{\ee}{\end{equation}}

\definecolor{viola}{RGB}{134,41,198}

\begin{document}

\title{Dark Matter from 't Hooft Anomaly Matching}

\author{Michele Redi}
\email{michele.redi@fi.infn.it}

\affiliation{
INFN Sezione di Firenze, Via G. Sansone 1, I-50019 Sesto Fiorentino, Italy
}
\affiliation{Department of Physics and Astronomy, University of Florence, Italy
}

\begin{abstract}
Light fermions that arise as composite states in confining gauge theories are stable due to fermion number conservation. 
This leads to Dark Matter candidates that behave similarly to elementary DM but where the cosmological stability arises automatically 
from the accidental symmetries of the theory. We present explicit models based on 't Hooft anomaly matching conditions where DM is charged under the SM.
For example an SU(N)  gauge theory with fermions in the adjoint rep and triplet of SU(2)$_L$ gives rise to DM made of multiple ``Wino'' triplets, 
while SO(N) gauge theories with $N_F=N-4$ flavors produce neutralino-like systems or SM quintuplets. Compositeness effects are crucial to determine the 
phenomenology.
\end{abstract}

\maketitle

\paragraph{\bf Introduction.}Despite many efforts the nature of Dark Matter (DM) continues to be a mystery. 
One possible clue is DM cosmological stability that suggests the existence of new accidental symmetries 
beyond baryon number of the Standard Model (SM). In the SM stability of proton and nuclei is an automatic consequence of the gauge structure of
the renormalizable lagrangian: given the quantum numbers of SM particles baryon number is an accidental symmetry of the theory that is only broken by dimension 6 operators or higher.
Certainly it would be desirable that stability of DM is also a robust feature of the theory, rather than being imposed by hand as often assumed in the literature. 
The request of DM accidental stability turns out to be very restrictive, demanding either new gauge symmetries \cite{Antipin:2015xia} or exotic SM representations \cite{Cirelli:2005uq}.
One of the simplest realizations of accidental DM is a baryon of a dark sector \cite{Antipin:2015xia}, see \cite{Kribs:2016cew} for a review.
%and {\cite{Buttazzo:2019iwr} for realizations with scalars.  
These scenarios  assume QCD-like dynamics, where a dark gauge group confines and spontaneously breaks global chiral symmetries of the fermionic lagrangian. 
In this letter we introduce a new class of composite DM models based on dynamics different from QCD leading to very different and exciting phenomenology. 
As argued by 't Hooft long ago \cite{tHooft:1979rat} strong coupling may result into color confinement without chiral symmetry breaking. 
When this phase is realized, examples of which have been proven in supersymmetric theories, massless composite fermions must exist in the chiral limit due to ``anomaly matching'': quantum anomalies 
of the symmetries of the full theory must be reproduced by the low energy degrees of freedom. If the elementary fermions are vectorial a small mass can be added leading to massive composite fermions, 
parametrically lighter than the confinement scale. As we will show the light composite fermions are excellent DM candidates.

\paragraph{\bf General structure.}
We will study the possibility that DM is a composite light fermion of a dark sector arising from 't Hooft anomaly matching.
Because of accidental fermion number conservation the lightest fermion is automatically stable at the renormalizable level,
leading to cosmological stability under very broad assumptions. We will focus here on  sectors with electro-weak charges 
that give rise to very predictive scenarios.

Differently from DM as a baryon of a dark sector, light composite fermions would behave for many aspects as elementary particles 
up to corrections $M/\Lambda$  where $M$ is the DM mass and $\Lambda$ the compositeness scale. For example if DM has SM  charges the annihilation cross-section that controls
the thermal abundance of DM  is the same of elementary SM multiplets to leading order.
This should come as no surprise, at low energies we can compute the pion production cross-section $\sigma(e^+ e^- \to \pi^+ \pi^-)$  using QED. 
If DM is produced thermally the abundance $\Omega\equiv \rho/\rho_c$ is then,
\begin{equation}
\Omega (M, \Lambda)= \Omega_{SM}(M)+ {\cal O} \left(\frac{M^2}{\Lambda^2}\right)
\end{equation}
The DM abundance is reproduced for $\Omega\approx 0.25$.
An important caveat to the almost elementary behaviour is provided by light composite singlets where the leading interactions depend instead on the compositeness scale.
As we will see the presence of singlets leads to strong constraints on $\Lambda$.

The main technical tool in the construction of explicit models is 't Hooft anomaly matching  conditions \cite{tHooft:1979rat} and recent generalizations \cite{Gaiotto:2017yup}. This allows one to determine gauge theories where confinement without chiral symmetry breaking is possible and the quantum numbers of the composite fermions. Our examples are non-supersymmetric theories where the composites can saturate anomalies but we cannot prove 
confinement without chiral symmetry breaking. Nevertheless we believe that these examples capture the essential features relevant for DM.
Confinement without chiral symmetry breaking can be established in supersymmetric theories, see \cite{Intriligator:1995au}.

\paragraph{\bf SU(N) with 3 adjoints.}

Our first example is an SU(N) gauge theory with adjoint fermions transforming as a triplet of SU(2)$_L$.
Recently it has been proposed in \cite{Poppitz:2019fnp} that SU(N) gauge theories with 3 Weyl adjoint fermions
may confine without chiral symmetry breaking. In the massless limit anomaly matching conditions can be most simply satisfied with $N^2-1$ Weyl fermions
that transform as triplets of the SU(3)  global symmetry.  The composite fermions are associated to the gauge invariant operators
\begin{equation}
O_n^i={\rm Tr}[G_{\mu\alpha_1}\dots G^{\alpha_n}_\nu \sigma^{\mu\nu} V^i]
\end{equation}
where $V^i\equiv V^{a,i} T^a$ is a triplet Weyl fermion in the adjoint of SU(N) and the trace is over adjoint indexes. 
The weak gauging of the global symmetry by the electro-weak interactions produces $N^2-1$ massless SU(2)$_L$ triplets. 
In order to avoid low scale Landau poles we will take here $N=3$.

Since the gauge theory is vector-like a mass $M_V$ can be added to the elementary fermions. The mass term breaks the SU(3) global symmetry to SO(3) allowing the composite fermions to also acquire a mass. Since the composite fermions are in a reducible representation of the global symmetry we expect their mass to be split by strong interactions, $M_i =c_i  M_V$.  It seems natural that fermions associated to  higher dimensional operators are heavier but we cannot prove it or estimate the splitting. We will assume that $c_i$'s are order one numbers in what follows.
In addition electro-weak interactions will further split different multiplet, $\Delta M_i/M_i \sim \alpha_2/(4\pi) \log \Lambda/M_i$ effectively renormalizing $c_i$. 
As in models with elementary triplets the charged components of each multiplet are split by $\Delta M= 165$ MeV.

Notably the underlying gauge theory is the same employed in Ref. \cite{Contino:2018crt} where theories with fermions in the adjoint 
of SU(N) were shown to produce an accidental DM candidate.  Assuming the standard pattern of symmetry breaking 
$SU(3)\to SO(3)$, the bound state made of adjoint fermions and gluons, the ``gluequark'', is a triplet of SU(2)$_L$ that is stable due fermion parity that is only violated by the dimension 6 operator, 
$V^{a,i} G_{\mu\nu}^a \sigma^{\mu\nu} L \sigma^i H$ suppressed by a scale $\Lambda_{\rm UV}$.  For $M_V< \Lambda$ the mass of the gluequark is expected to be of order $\Lambda$. 

If confinement produces massless chiral fermions the discrete symmetry still guarantees the stability of the lightest fermion, but the DM phenomenology is entirely different.  First, the gluequarks have multiplicity $N^2-1$. The lifetime induced by the parity violating operator becomes,
\begin{equation}
\frac 1 {\tau_{\rm DM}} \sim \frac {M_{\rm DM}\Lambda^4}{8\pi\Lambda_{\rm UV}^4} \sim \frac{10^{-28}}{{\rm s}}\left(\frac {M_{\rm Pl}}{\Lambda_{\rm UV}}\right)^4 \left(\frac {\Lambda}{100\, {\rm TeV}}\right)^4 \left(\frac {M_{\rm DM}}{ {\rm TeV}}\right)
\end{equation}
which is parametrically larger than in the composite regime by a factor $M_{\rm DM}/\Lambda$. 
Moreover the gluequarks now behave at low energies as elementary electro-weak triplets. Since the DM thermal abundance 
is determined by electro-weak interactions the critical mass is in the TeV range, rather than ${\cal O}(100)$ TeV as in \cite{Contino:2018crt,Antipin:2015xia}. 
If dim 6 operators are suppressed by the Planck scale cosmological stability demands $\Lambda< 1000$ TeV. 
It is also conceivable that quantum gravity  effects are further suppressed so that larger values of $\Lambda$ become allowed.

In the  effective theory below $\Lambda$ all the triplets are stable at the renormalizable level. Denoting with $\lambda_3^i$ and $\lambda_{3'}^i$ the
triplets of mass $M_3$ and $M_{3'}$ respectively, the heavier one can decay through dimension 7 operators suppressed by the compositeness scale \footnote{4-Fermi operators could also allow the decay if kinematically allowed. This however requires $M_{3'}>3 M_3$. We will not consider this possibility here.},
%\begin{equation}
%\frac {\alpha_2^2}{(4\pi)^2 \Lambda^3} \lambda_{3}^a  \lambda_{3'}^a W_{\mu\nu}^b W^{b\,\mu\nu}
%\end{equation}
\begin{equation}
\frac {\alpha_2 \alpha_*}{\Lambda^3} \lambda_{3}^i  \lambda_{3'}^i W_{\mu\nu}^j W^{j\,\mu\nu}
\end{equation}
where $\alpha_*\sim 1$ is the strong dynamics coupling and we included a loop factor in the NDA estimate as for dipole operators \cite{Giudice:2007fh}. For $\Delta M > 2 M_W$ the lifetime is,
\begin{equation}
\frac 1 {\tau_{3'}} \sim \frac {\alpha_2^2 \alpha_*^2} {192\pi^3}  \frac {M_{3'}^7}{\Lambda^6}\sim \frac{10^{-28}}{{\rm s}} \left(\frac {M_{3'}}{\rm TeV}\right)^7 \left(\frac {10^{11}\, \rm GeV} {\Lambda}\right)^6
\label{3decay}
\end{equation}
If decay through emission of $W$ gauge bosons is kinematically forbidden the main decay is through photon emission with a  rate obtained replacing $\alpha_2\to \alpha_{em}$.

\begin{figure}[t!]
\centering
\includegraphics[width=.98\linewidth]{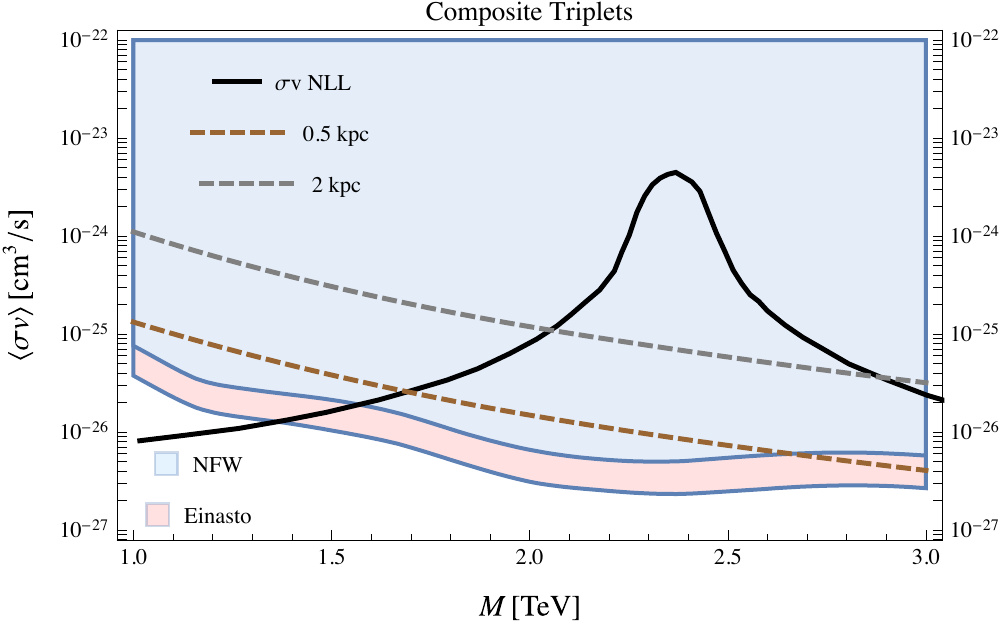}
\vspace{-0.2cm}
\caption{
\label{fig:triplet-indirect}
\it Indirect detection constraints for DM triplets annihilation to photons in the galactic center. The black line is the NLL annihilation cross-section \cite{Rinchiuso:2018ajn}.
Red and blue regions are HESS exclusion limits \cite{Abdallah:2018qtu}  respectively for Einasto and NFW DM profiles, rescaled to account 
for the different numerical abundance. Also shown in dashed brown and green are the projections for DM with core radii 0.5 and 2 kpc extracted from \cite{Rinchiuso:2018ajn}.
If the multiplets are degenerate bounds 8 times stronger apply.}
\end{figure}

Different regimes are realised depending on the scale of compositeness:\\
For $\Lambda \gtrsim 10^{11}$ GeV all the triplets could be cosmologically stable if dim 6 parity violating operators are suppressed,
giving rise to multi-component DM made of neutral components of triplets. The thermal abundance is determined by annihilation into SM particles as for elementary Winos.
The tree level s-wave annihilation cross-section of a triplet of mass $M_i$ reads,
\begin{equation}
\langle \sigma_i v_{\rm rel} \rangle = \frac {37}{12}\frac {\pi \alpha_2^2}{M_i^2}
\end{equation}
which should be multiplied by the appropriate Sommerfeld enhancement factors, see \cite{Baumgart:2018yed} for precision computations. The abundance of each component is inversely proportional to  its annihilation cross-section. Since a thermal Wino has a mass 2.9 TeV  the critical abundance of DM is now obtained for $\sum_i^{8} M_i^2\approx (3\, \rm TeV)^2$,  if DM is a thermal relic.  Contrary to \cite{Buen-Abad:2015ova} DM triplets are not degenerate and this leads to distinctive features. The numerical thermal abundance of each DM component is $n_i \propto M_i$ so that heaviest species is the most abundant. The astrophysical signals will depend on the distribution  of masses. In particular each DM component  can annihilate into photons giving rise to monochromatic photons with a rate rescaled by $(M_i/3\,{\rm TeV})^4$ compared to a Wino of the same mass comprising the totality of DM. For a single triplet in fact the rate from the galactic center 
is in strong tension with HESS constraints \cite{Abdallah:2018qtu}, under the assumption of Einasto or NFW DM profile. 
Necessarily in our case $M_i< 3$ TeV weakening the constraints. The details depend on the value $c_i$. If the masses are almost degenerate the relic abundance is reproduced for $M_i\approx 1$ TeV and the astrophysical signals are reduced by a factor 8 compared to a Wino. If the multiplets are split then each component would annihilate producing monochromatic photon of energy $E_\gamma\approx M_i$ with approximately constant intensity. Lines split by at least 10\% can be resolved by HESS \cite{Rinchiuso:2018ajn} or CTA \cite{Rinchiuso:2020skh}. Resolution up to percent level could be reached by future experiments such as HERD \cite{Fusco:2019zna}. Current bounds from HESS are shown in Fig. \ref{fig:triplet-indirect}. 

For $10^9\, {\rm GeV}\lesssim \Lambda \lesssim 10^{11}\, {\rm GeV}$ the lifetime of heavier states is longer than the age of the universe but their decay can produce astrophysical signals. 
For $10^5\,{\rm GeV} \lesssim \Lambda \lesssim 10^{9}$ GeV heavier triplets decay after freeze-out into the lightest component. 
This slightly offsets the DM abundance because each decay yields exactly 1 DM particle.
The DM abundance is reproduced for $M_1\sum_{i=1}^{8} M_i\approx (3\, \rm TeV)^2$. 
For $\Lambda\sim 10^6$ GeV the lifetime is around 1s potentially affecting Big Bang Nucleosynthesis.
Similarly if the decay takes place during recombination bounds from the energy injected in the SM plasma could arise.

Finally for $\Lambda \lesssim 100 $ TeV the decay rate of heavier states is faster than Hubble at freeze-out. 
Therefore different species remain in equilbrium co-annihilating with each other \cite{Griest:1990kh}. 
If the splitting is significant the abundance becomes the one of an elementary Wino while for small splittings
the abundance is modified approaching 8 times the Wino value.

Triplets in the TeV range can be searched at colliders through disappearing tracks. A single triplet with mass 3 TeV is borderline at a 100 TeV collider \cite{Low:2014cba}. 
In the composite scenario the multiplicity and possibly the lower mass reproducing thermal abundance makes DM within reach of future colliders.

\paragraph{\bf SO(N) with N-4 fundamentals.}

Our second example is an SO(N) gauge theory with $N-4$ fundamental fermions (F) and 1 adjoint (A).
In the massless limit the anomaly free global symmetry is $SU(N-4)\times U(1)$. Anomalies can be matched by the massless fermion,
\begin{equation}
\Psi^{ij}= F^i_\alpha A_{\alpha\beta} F_\beta^j
\end{equation}
in the symmetric rep of $SU(N-4)$. The quantum numbers of the preons and  composite fermions are thus given by,
\begin{equation}
\begin{tabular}{c|ccc}
\hbox{Name}&   $SO(N)$    & $SU(N-4)$   & U(1)\\  \hline
F & $\yng(1)$  &   $\yng(1)$ &  $-\frac {N-2}{N-4}$   \\ 
A & $\yng(1,1)$ & 1  & 1   \\  \hline 
$\Psi$ &1 & $\yng(2)$ & $-\frac {N}{N-4}$
\end{tabular}
\label{table:boundstates}
\end{equation}
The fermion content is identical to $N=1$ supersymmetric SO(N) Yang-Mills with $N-4$ flavors that is known to confine without chiral symmetry breaking
but the composite fermions are different in our case.

Adding masses $M_{F,A}$ for the elementary fermions breaks the global symmetries allowing to generate a mass 
for the composites. The spurionic quantum numbers of mass deformation follow directly from eq. (\ref{table:boundstates}). 
In order to match the global symmetries, to leading order the mass term for the composite fermions must be proportional to $M_F^2 M_A$.
Two structures can be written down compatibly with the symmetries,
\begin{equation}
{\cal L}_M= a_1 M_A {\rm Tr}[M_F \Psi]^2 + a_2 M_A |{\rm Tr}M_F  \Psi M_F \Psi| + h.c.
\label{c1c2}
\end{equation} 
where $M_F$ is an $(N-4)\times (N-4)$ matrix and $a_{1,2}$ are determined by the dynamics. 

This setup allows one to construct theories with Majorana DM (Sp(N) gauge theories could be used
to produce pseudo-real DM, see \cite{Gertov:2019yqo} for a related construction). 

Simplest examples are as follows:\\

$\bullet\, N_F=3:$ The elementary fermions transform as a triplet of SU(2)$_L$. The composite fermions are then,
\begin{equation}
\yng(2)=6=1+5
\end{equation}
giving rise to a singlet $\lambda_1$ and an SU(2)$_L$ quintuplet $\lambda_5^{ij}$. 
%The smallest parity violating operator has a dimension 6 mixing the singlet with right-handed neutrinos.
Note that since the only parameter is the elementary triplet mass $M_V$ the ratio of $M_1$ and $M_5$ is fixed in terms of $a_{1,2}$.
The existence of a singlet changes  the abundance of DM compared to the elementary quintuplet scenario \cite{Cirelli:2005uq}. 
All the singlet interactions are controlled by the compositeness scale through higher dimension operators such as,
\begin{equation}
\frac {\alpha_2 \alpha_*}{\Lambda^3} \lambda_5^{ij}  \lambda_1 W_{\mu\nu}^i W^{j\,\mu\nu}\,,~~~~\frac {g_*^2}{\Lambda^2} \lambda_1^2 \lambda_5^2\,, ~~~~\frac {g_2^2}{\Lambda^2} \lambda_1^2 f_{SM}^2
\end{equation}
The first operator allows the heavier state to decay with a lifetime (\ref{3decay}).
The last two operators (induced for example by composite spin-1 tree level exchange) allow the singlet to annihilate into quintuplets or SM fermions $f_{SM}$ with a  cross-section $\sigma v \sim \pi \alpha_{2}^2 M_1^2/\Lambda^4$.

\begin{figure}[t!]
\centering
\includegraphics[width=.98\linewidth]{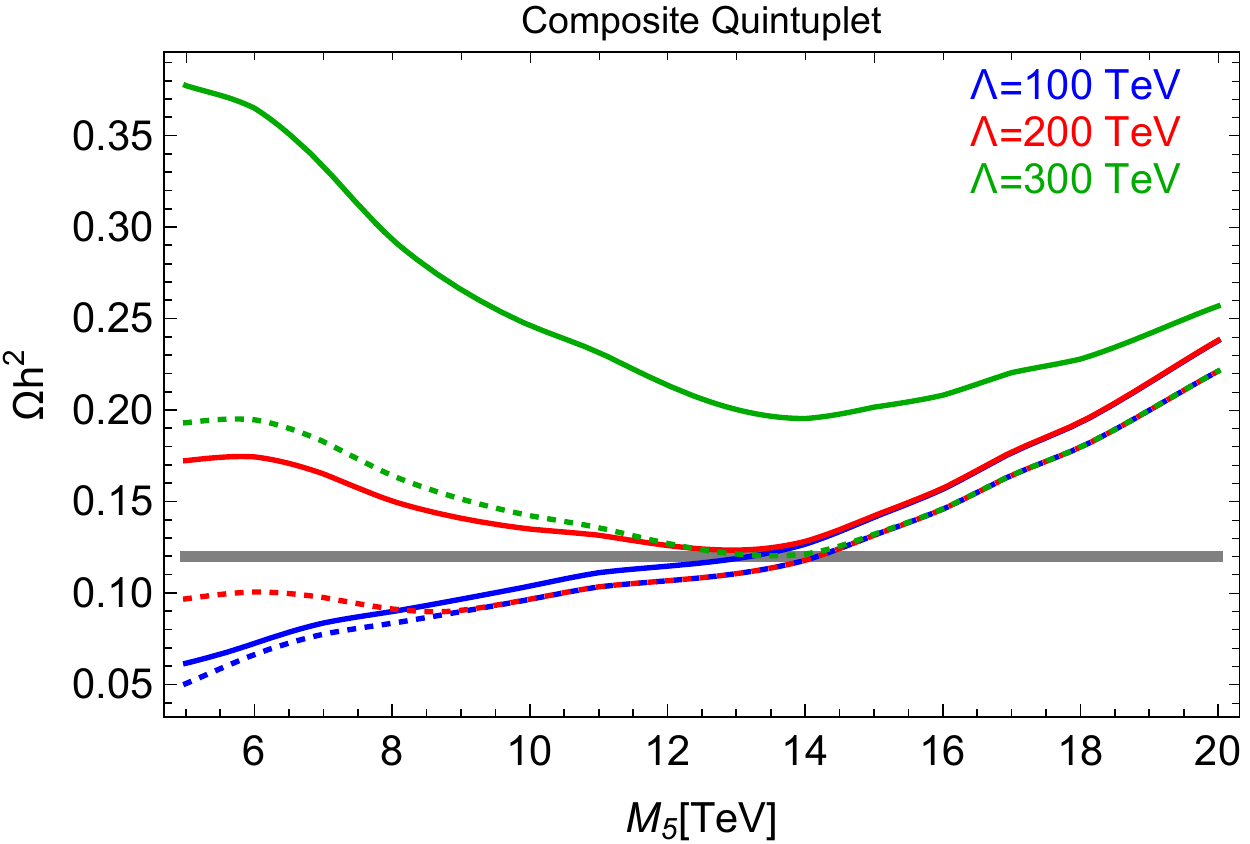}
\vspace{-0.2cm}
\caption{
\label{fig:SO}
\it Relic abundance estimate of DM made of SU(2)$_L$ quintuplets for different values of $\Lambda$. 
Solid and dashed lines correspond to $\Delta M/M=0.02$ and $\Delta M/M=0.1$ respectively.}
\end{figure}

The cosmological abundance depends strongly on the compositeness scale.
In the mass range considered, for $\Lambda \lesssim  100$ TeV, singlet and quintuplet are in equilibrium at least until freeze-out due to the fast decays of the heavier state. The abundance of DM is thus determined by co-annihilation of the two states. If $M_5<M_1$ the result is similar to the elementary quintuplet  where the thermal abundance of DM is reproduced for $M_{\rm DM}\approx$ 14 TeV \cite{Mitridate:2017izz}. For larger values of $\Lambda$ the abundance is modified by late time decays of the singlet. The thermal abundance can be reproduced for $M< 14$ TeV changing the experimental  predictions, see \cite{Mahbubani:2020knq} for a recent discussion. In Fig.  \ref{fig:SO} we show the relic abundance of quintuplets  obtained by solving the coupled singlet-quintuplet Boltzmann equations for different splittings and $\Lambda$. 

If the singlet is the lightest state its small annihilation cross-section into SM generates a large abundance of DM. 
Reproducing the critical DM abundance requires either a very small scale $\Lambda$ or a small splitting with the quintuplet 
so that the singlet co-annihilates strongly.

$\bullet\, N_F=5:$ The elementary fermions transform as $L+\bar{L}+N$. The composite fermions are given by,
\begin{equation}
\yng(2)=15=2\times 1_0 + 3_{\pm 1}+ 3_0+ 2_{\pm \frac 1 2}
\end{equation}
The resulting system is similar to neutralinos in supersymmetry with extra singlet and triplets with hypercharge. 
The elementary lagrangian contains,
\begin{equation}
-{\cal L}_M= M_L L \bar{L}+ M_N N^2 + y L H N + \tilde{y} \bar{L} \tilde{H} N
\end{equation}
The low energy lagrangian is found by treating the $M_F$ in (\ref{c1c2}) as a function of the electro-weak VEV.
This generates effective Yukawa interactions between doublet and singlets and doublets and triplets.

After electro-weak symmetry breaking different multiplets mix producing a set of neutral Majorana particles the lightest of which is identified with DM.
This allows to avoid strong direct detection constraints from tree level exchange of $Z$ boson. 
Experimental signatures are analogous to neutralinos including direct detection signals and CP violating effects, see \cite{Mahbubani:2005pt,Calibbi:2015nha}.
Let us note that the elementary lagrangian contains 4 (complex) parameters so that the composite lagrangian parameters are restricted.

In the limit $M_N \gg M_L$ the system reduces effectively to $N_F=4$ with fermions $L+\bar{L}$. The subset of composite fermions
are
\begin{equation}
\yng(2)=10= 1_0 + 3_{\pm 1}+ 3_0.
\end{equation}
The mass term $L\bar{L}$ breaks $SU(4)\to SO(4)\simeq SU(2)\times SU(2)$ under which the triplets transform as $(3,3)$. 
The triplets are thus degenerate up to electro-weak symmetry breaking effects. This implies that the heavier triplets are very long lived. 
If the triplets with hypercharge are cosmologically stable direct detection constraints from tree level Z-exchange would 
exclude the model. This potential danger would be eliminated introducing a small mixing with singlet such that the Majorana states are split by 100 KeV or more.

\paragraph{\bf Outlook.}

In this letter we proposed a new realization of accidental DM based on confining gauge theories. 
If the theory confines without chiral symmetry breaking, the lightest composite fermion required by 
't Hooft anomaly matching conditions is automatically stable, providing a promising DM candidate. 
For simplicity we focused on the possibility that the elementary  fermions have electro-weak charges realizing a WIMP-like scenario.

While DM is intrinsically composite it behaves as elementary up to corrections $M/\Lambda$, for what concerns annihilation and 
low energy processes.  Explicit models often feature several states cosmologically stable up to compositeness effects, leading to multi-component DM or 
effects in the thermal abundance of DM. Compositeness has particularly dramatic effects for SM singlets.
They can decay through high dimension operators suppressed by the compositeness scale and their interactions are completely controlled by $\Lambda$.
In particular singlets tend to overproduce DM if $\Lambda$ is large.

For high compositeness scale DM can be made of several SM multiplets leading to a distinctive phenomenology within the reach of future experiments.
Among potential signals are multiple photon lines in the TeV range that could be resolved by future experiments. High luminosity LHC or 100 TeV 
collider could also  discover the new states more easily than elementary multiplets in light of multiplicity and smaller mass.

Many extensions of this work can be envisaged.
In the examples presented we focused on Majorana DM. Dirac DM could be constructed in theories that preserve dark baryon number, most simply in SU(N) gauge theories with fundamentals.  
We leave a thorough classification of possible models and the detailed study of the phenomenology to future work. 
In non-supersymmetric gauge theories currently one cannot prove that the dynamics realizes confinement without chiral symmetry breaking.  
In order to establish the dynamics lattice QCD methods could be used. This is similar to the search of the conformal window in confining gauge theories \cite{DeGrand:2015zxa}
with the notable difference that only certain fermionic correlators would display a power law behaviour at long distances.
In supersymmetric gauge theories very well known examples exist of confinement without chiral symmetry breaking \cite{Intriligator:1995au} . 
Using the results of \cite{ArkaniHamed:1998wc} supersymmetry breaking effects can be included in a controllable way allowing to construct explicit models within theoretical control.
Finally the addition of scalars is expected to lead to many other scenarios with composite light fermions \cite{Dimopoulos:1980hn} that can be suitable for DM.

%%%%%%%%%%%%%%%

\medskip
\paragraph{ \it Acknowledgements.}\,\,
{\small This work is supported by MIUR grant PRIN 2017FMJFMW. I would like to thank Roberto Contino, Ken Konishi and Eric Poppitz for useful discussions.}

%%%%%%%%%%%%%%%%%%%%%%%%%%%%%%%%%%%%%%%%%%%%%%

\bibliography{biblio}

\end{document}